\documentstyle[aps,manuscript,epsfig]{revtex}
\begin{document}
%-----------------------------------
%   Title
%-----------------------------------
\title{Forced Topological Nontrivial Field Configurations}
%-----------------------------------
%   Authors
%-----------------------------------
\author{V.G. Kiselev$^{a,b}$, Ya.\ M. Shnir$^{a,c}$}
%-----------------------------------
%   Address
%-----------------------------------
\address{
$^a$Institute of Physics,
Academy of Sciences of Belarus, Minsk, Belarus \\
$^b$Institute of Medicine, Research Centre J{\"u}lich, 
 52425 J{\"u}lich, Germany\\
$^c$DAMTP, University of Cambridge, Silver Street, Cambridge CB3 9EW, UK}
\maketitle 

\thispagestyle{empty}

\bigskip

%-----------------------------------
%   Abstract
%-----------------------------------
\begin{abstract}

The motion of a one-dimensional kink and  its energy losses are  
considered as a model 
of interaction of nontrivial topological field configurations with  
external 
fields. The approach is based on the calculation of the zero modes 
excitation  probability in the external field. 
We study in the same way the interaction of the t'Hooft--Polyakov monopole 
with weak 
external fields. The basic idea is to treat the excitation of a monopole 
zero mode as the monopole displacement.  The excitation is found 
perturbatively.  As an example we consider the interaction of the 
t'Hooft-Polyakov 
monopole with an external uniform magnetic field. 
\end{abstract}

 
\section{Introduction} In the approximately 20 years, which have 
passed from the time of understanding the role  the topological 
nontrivial field configurations, such as kink \cite{AdDash}, vortex 
\cite{NielsOles} and monopole \cite{Hooft},\cite{Pol} are playing in different 
branches of physics,  considerable progress has been achieved.  There 
is a large number of papers devoted to both the classical and quantum 
aspects of the problem. 
 
Nevertheless, there  are still  some open problems connected with the 
description of  the interaction between these objects as well as their 
interaction with the external fields.  
For example, by calculation of the static force between two monopoles 
\cite{Manton-1} as well as by  description of monopole-monopole 
scattering \cite{Manton-2}, only the Bogomolny-Prasad-Sommerfield (BPS) 
limit \cite{BPS} was considered. The same BPS monopoles were studied by 
calculation of light scattering by a monopole \cite{BakLee}. A typical 
feature of these calculations is their classical character. The main 
reason for using  the BPS approximation is that there are analytical 
solutions both for the monopole structure functions and its mass as 
well as a clear mathematical description of this classical field 
configuration. On the other hand, for the BPS monopoles one can 
construct a moduli space of exact multimonopoles solution and describe 
their low-energy scattering by geodesic motion on this space 
\cite{Manton-3}. 

But, as  was shown in Refs. \cite{KisSel} and \cite{Zar}, if one takes 
into account the quantum fluctuation on the  monopole background field, 
there is no consistent BPS limit because when one reaches it the 
quantum correction turns out to be increasing, causing the limiting 
transition to be impossible. Another assumption in consideration of the 
interaction of the BPS monopole with an external homogeneous magnetic  
field is just using {\it ad hoc} of the ansatz for the 't Hooft-Polyakov 
field configuration moving with a constant acceleration  
\cite{Manton-1},\cite{BakLee}.  
 
In the present paper we develop the  
consistent perturbative consideration of the problem of the interaction  
of nontrivial topological field 
configurations with external fields in a simple example of the kink 
acceleration and radiation in (1+1)-dimensional $\lambda \phi^4$ 
theory. The same approach applied to the consistent description 
of a more complicated case of the interaction between  
the 't Hooft-Polyakov monopole solution  and  
an external  weak uniform  magnetic field.  
 
\section{Acceleration and energy losses of two-dimensional kink} 
 
We start from the Lagrangian of a two-dimensional model  
\begin{equation}                                    \label{Lagrang} 
L = \frac{1}{2} {\dot {\phi}}^2 - \frac{1}{2}{{\phi}'}^2 - \frac{\lambda} 
{4}\left(\phi ^2 - \frac{m^2}{\lambda}\right)^2 - \varepsilon   
\frac {m^3}{\sqrt \lambda} \phi,   
\end{equation} 
where the dimensionless parameter $\varepsilon \ll 1$. The potential of  
the model  
\begin{equation}                                    \label{Potential} 
V[\phi] = \frac{\lambda}{4} 
\left(\phi ^2 - \frac{m^2}{\lambda}\right)^2 + \varepsilon  \frac {m^3} 
{\sqrt \lambda} \phi 
\end{equation} 
has two minima (see fig.1).  
So, the Lagrangian (\ref{Lagrang})  
corresponds to the so-called thin-wall approximation  
\cite{KisSel-2}, \cite{Vol}  
of the well-known problem of the  
spontaneous vacuum decay \cite{VolKobOk}, \cite{LifKag}.  
Let us recall only that at $\varepsilon = 0$ the two vacua are degenerate and  
there is a topological nontrivial kink solution $\phi _0$ that interpolates  
between these vacua (see e.g. \cite{Radj}). It can move uniformly along the  
$x$ direction that is connected with zero mode presence. 
If $\varepsilon \neq 0$ one can consider the problem as a kink under an   
external weak force which corresponds to the linear term in (\ref{Potential}).  
 
Note, that in the classical case this system has a very simple   
interpretation in the solid state physics: it is 
the continuum representation of  
the model of a structurally unstable ion lattice, 
having a double-well local potential and nearest-neighbor coupling \cite{KS}. 
The kink configuration in this picture corresponds to the domain wall and the  
continuum modes ($\eta _k $) are just phonons. 
 
Let us consider the evolution of the kink after the metastable vacuum decay. 
In order to solve the field equation corresponding to the Lagrangian  
(\ref{Lagrang}) 
\begin{equation}                                    \label{field-equ} 
{\ddot {\phi}} - {\phi}'' - m^2 \phi + \lambda \phi ^3 + \varepsilon \frac{m^3}
{\sqrt \lambda} = 0 
\end{equation} 
we can use an expansion in powers of $\varepsilon$:                             
$\phi = \phi_0 + \varepsilon \phi_1 + \varepsilon ^2 \phi _2 + \dots $.  
 
In general, for calculation of the corrections to the kink field  in  
 $n$-th order of $\varepsilon$ we have an equation \cite{KisKob}: 
\begin{equation}               \label{n-equ} 
\biggl(\frac {d^2}{d t^2} + D^2\biggr) 
 \phi _n +  F(\phi_{n-1}, \dots \phi _0 ) = 0, 
\end{equation} 
where the operator $D^2$ is 
\begin{equation}                   \label{second-der} 
D^2 = - \frac {d^2}{d x^2} - m^2 + 3m^2 \tanh^2~ \frac {mx}{\sqrt 2} = -m^2 \left(
\frac{1}{2} \frac{d^2}{d z^2} + 1 - 3 \tanh^2 z \right),
\end{equation} 
$z =  {mx}/{\sqrt 2}$ 
and $F(\phi_{n-1}, \dots \phi _0 )$ is a function of all 
 low order corrections  $\phi _k$. $k < n$. Note, that parity of  
$F(\phi_{n-1}, \dots \phi _0 )$, as well as $\phi _n$ are  
interchanging from one  
order of  $\varepsilon$ to another.  Thus, one can pick out the asymptotic of  
the $n$-th order correction by definition 
$$\phi_{2n-1}(z) =  B_{2n-1} + \chi_{2n-1}(z); \qquad \phi_{2n}(z) =  
B_{2n} \tanh z  + \chi_{2n}(z), 
$$ 
where $B_{k} = ~{\rm constant}~$. The boundary condition is that the  
functions $\chi_{k}(z)$ tend to zero at $z \to \infty $. 
 
Thus,  
the zero-order approximation gives the classical equation  
\begin{equation}                                             \label{zero-equ} 
{\ddot {\phi _0}} - {\phi _0}'' - m^2 \phi _0 + \lambda \phi _0 ^3 = 0, 
\end{equation} 
with the above-mentioned kink solution \cite{AdDash}  
\begin{equation}                   \label{kink} 
\phi _0 = \frac{m}{\sqrt {\lambda}} \tanh z
%~{\rm {th}}~  
\end{equation} 
 
The first order corrections to the  solution (\ref{kink}) can be obtained from 
the next equation 
\begin{equation}                   \label{first-equ} 
\frac {d^2}{d t^2} \phi _1 + D^2 \phi _1 +  \frac {m^3}{\sqrt \lambda}  = 0, 
\end{equation} 
where $D^2$ is the operator 
(\ref{second-der}). 

 In order to find the corrections to the kink solution we can use the 
expansion  
of  $\phi _1$ on the normalizable eigenfunctions 
$\eta _n (z)$ of the operator $D^2$ which describe the scalar field  
fluctuations on the kink background, i.e. one can write 
\begin{equation}                            \label{expan} 
\phi _1 = \sum _{n=0}^{\infty} C_n(t) \eta _n (z), 
\end{equation} 
where the solutions of the eigenvalue problem  
 $D^2\eta _n (z) = \omega _n^2\eta _n (z)$ are (see e.g., \cite{Radj}) 
\begin{eqnarray}                                \label{modes} 
\eta _0 (z) = \frac{1}{\cosh^2~z};\qquad 
\eta _1 (z) = \frac{\sinh z}{\cosh^2~z}; 
\nonumber\\[4pt] 
\eta _k (z) = e^{ikz} (3 \tanh ^2~z - 3ik \tanh z -1 - k^2). 
\end{eqnarray}  
The corresponding eigenvalues are 
\begin{equation} 
\omega _0^2 = 0; \qquad  \omega _1^2 = \frac{3}{2} m^2; \qquad 
\omega _k^2 = m^2 \biggl( 2 + \frac{k^2}{2} \biggr). 
\end{equation} 
So, there is a zero mode ($\eta _0 $), which corresponds to kink translation, 
a vibrational mode ($\eta _1 $), connected with the time-dependent deformation  
of the kink profile, and continuum modes ($\eta _k $) which in quantum theory  
correspond 
to scalar particle excitations on the kink background.  
These functions form  
a complete set which spans the space of any function of $z$. 
The corresponding orthogonality relations are 
\begin{eqnarray}                 \label{ort} 
\int\limits_{- \infty}^{\infty} \eta _0^2 dz = \frac{4}{3};
\qquad  \int\limits_{- \infty}^{\infty} \eta _1^2 dz = \frac{2}{3}; 
\nonumber  \\ 
\int\limits_{- \infty}^{\infty} \eta _k ^{*}\eta _{k'} dz = 
{2 \pi} (1 + k^2)(4 + k^2) \delta (k - k'). 
\end{eqnarray}  
 
If we substitute the expansion (\ref{expan}) into eq.(\ref{first-equ}) 
we obtain 
\begin{equation}                   \label{first-equ-exp} 
\sum _{n=0}^{\infty}  
\biggl({\ddot C}_n(t) + \omega _n ^2 C_n(t)\biggr) \eta _n (z)    + 
 \frac {m^3}{\sqrt \lambda}  = 0 
\end{equation} 
 
Using the orthogonality relations (\ref{ort}) one can make a projection of   
eq.(\ref{first-equ-exp}) onto the modes $\eta _n (z)$. The projection onto  
the zero modes gives the equation (here we take into account that  
$\int\limits_{- \infty}^{\infty} \eta _0 dz = 2 $): 
\begin{equation}                   \label{first-equ-zero} 
\frac{4}{3}{\ddot C}_0 + \frac{2 m^3 }{\sqrt {\lambda}} = 0 
\end{equation} 
with the solution 
$$ 
C_0 = - \frac{3m^3}{4 \sqrt {\lambda} } t^2 + V_0 t + x_0. 
$$  
It means that the correction to the kink solution due to zero mode  
excitation is (here we suppose that $V_0 = x_0 = 0$): 
\begin{equation}            \label{kink-corr-zero} 
\phi =         
\phi _0 (z) + \varepsilon C_0(t) \eta_0 (z) = \frac{m}{\sqrt {\lambda}}  
\tanh z - \varepsilon 
\frac{3 m^3 }{4 {\sqrt {\lambda}} } \frac {t^2}{\cosh ^2~z} 
= \phi _0 (z + \delta z^{(1)})  
\end{equation} 
where the shift of the kink to the first order is given by  
$$ \delta z^{(1)} =  -\varepsilon \frac {3 m^2 }{4 } t^2, \quad ~{\rm or} ~ \quad
\delta x^{(1)} = -\varepsilon \frac {3 m }{2 {\sqrt 2} } t^2. 
$$ The meaning of this correction is quite obvious: because the external  
force  $F$ we introduced in (\ref{Lagrang}) in the first order is  
(here $E$ is the energy density) 
$$F = -\int\limits_{-\infty}^{\infty}dx \frac{dE}{dx} =  -\int\limits_{-\infty} 
^{\infty} dx ~\frac {d}{dx}\left[  
\frac{1}{2} {\dot {\phi}}^2 + \frac{1}{2}{{\phi}'}^2 + \frac{\lambda}{4} 
\left(\phi ^2 - \frac{m^2}{\lambda}\right)^2 - \varepsilon  \frac {m^3}{\sqrt  
\lambda} \phi_0 \right] 
$$ 
$$ 
= -\frac {d M}{dx} -  \varepsilon \int\limits_{-\infty}^{\infty} dx ~\frac {d} 
{dx} \frac {m^3}{\sqrt \lambda} \phi_0  
= -2 \varepsilon  \frac {m^4 }{\lambda} 
$$  
where the kink energy ${\cal E}$ or its  classical mass $M$ is 
\begin{equation}            \label{kink-energy} 
{\cal E}  \equiv M = \int\limits_{-\infty}^{\infty} dx \left[  
\frac{1}{2} {\dot {\phi}}^2 + \frac{1}{2}{{\phi}'}^2 + \frac{\lambda}{4} 
\left(\phi ^2 - \frac{m^2}{\lambda}\right)^2 + \varepsilon  \frac {m^3}{\sqrt  
\lambda} \phi \right]= \frac {2 {\sqrt 2} m^3 } 
{ 3 \lambda} 
\end{equation} 
we see that really the acceleration of the kink is given by the relation 
$$ 
a = -\varepsilon \frac {3 m} {\sqrt 2} = \frac {F}{M} 
$$ 
that is exactly the Newton formula. 
 
The meaning of other corrections can be found in the same  
way. After the  
projection of eq.(\ref{first-equ-exp}) onto the vibrational mode 
$\eta _1 (z)$ we have  
\begin{equation}            \label{first-equ-one} 
{\ddot C}_1 + \omega _1^2 C_1 = 0, ~~{\rm i.e.}~~~C_1 = ~{\rm Constant}~  
e^{i\omega _1 t} 
\end{equation} 
i.e. there is no interaction between this mode and the external force.

The solution of eq. (\ref{first-equ-one}) is defined up to an arbitrary  
constant that, from dimensional arguments, can be written as 
$\frac{m}{\sqrt \lambda} a_1$, where $a_1$ is a dimensionless parameter 
fixed by  
the initial conditions at $t = 0$. In the framework of the classical theory it  
corresponds  
to the value of amplitude of the first vibrational mode.         
 
As for the k-th mode belonging to the continuum one obtains  
\begin{equation}            \label{first-equ-k} 
{\ddot C}_k + \omega _k^2 C_k + \frac { m^3 \int 
\eta _k (z) dz }{2 {\sqrt \lambda }\pi (1 + k^2)(4+ k^2)} = 0. 
\end{equation} 
 
Calculation of the integral here gives 
\begin{equation}  
\int\limits_{-\infty}^{\infty} dz~ \eta _k (z) = 2 \pi (2 - k^2) \delta (k) 
\end{equation} 
and we have \begin{equation}  
{\ddot C}_k + \omega _k^2 C_k + \frac { m^3 (2 - k^2) 
\delta (k) }{{\sqrt \lambda}  (1 + k^2)(4+ k^2)} 
= 0 
\end{equation} 
In case of the lowest mode of the continuum ($k_0 = 0$) 
it is just the equation for an oscillator in external field with 
the solution 
\begin{equation}               \label{first-equ-k'} 
C_{k_0} = e^{i{{\omega }_{k_0}} t} - \frac { m }{4 {\sqrt \lambda} } \equiv  
{\widetilde C}_0 - \frac  
{ m}{4 {\sqrt \lambda} } 
\end{equation} 
For all other continuum modes with $k\neq 0$ we have the trivial 
oscillator equation 
\begin{equation}               \label{first-equ-all} 
{\ddot C}_k + \omega _k^2 C_k = 0, ~~{\rm i.e.}~~~ C_k = ~{\rm Constant}~\cdot 
 e^{i\omega _k t}. 
\end{equation} 
Using the above mentioned arguments one can write the arbitrary  
constants as $\frac{m}{\sqrt \lambda} a_k$, where the parametrs  
$a_k$ are fixed by the initial   
conditions. 

Thus, collecting the contributions from all modes of  
excitation (\ref{first-equ-zero}), (\ref{first-equ-one}), (\ref{first-equ-k'}) 
 and (\ref{first-equ-all}), we find the first order correction to the kink  
configuration: 
\begin{equation}               \label{phi-one}             
\phi _1 = \frac {m}{\sqrt\lambda} \left\{ - \frac {3}{4} m^2 t^2 
{\eta }_0  - \frac {1}{4 } {\eta }_{k{_0}}   
+ a_1 e^{i \omega _1 t} \frac {\sinh z}{\cosh^2 z} +  
\sum _{k=0}^{\infty} a_k {\widetilde C}_k(t) \eta_k (z)\right\}  
\end{equation} 
where ${\widetilde C}_k(t) = e^{i \omega _k t}$ and ${\eta }_{k{_0}} =  
3 \tanh^2 z - 1 $. 
The last two terms in this expression correspond to the fluctuation corrections 
to the  
kink solution and can be excluded if we take the initial condition at 
$t = 0$ as 
$a_1 = 0, a_k = 0$ for all k. 
  
The first term, connected with the zero mode contribution,  
describes the motion of the kink with a constant acceleration, as  mentioned  
above. The meaning of the second term can be clarified if one considers the  
corresponding correction in the asymptotic region ($z \to \pm \infty $),  
where we have  (up to fluctuation corrections)   
\begin{equation}               \label{phi-vac}        
\phi (\pm \infty ) =  \frac {m}{\sqrt{\lambda} }\left(\pm 1 - \frac 
{\varepsilon }{2} + {\cal O}(\varepsilon^2)\right). 
\end{equation} 
Indeed, the potential (\ref{Potential}) has the minima at   
$\phi = \phi (\pm \infty )$ given by eq.(\ref{phi-vac}). 
Thus, this term corresponds to a shift of the vacuum value of the scalar  
field (see figs.~1,2). 
 
The expression (\ref{phi-one})  allows to calculate the first order 
corrections  
to the kink energy ${\cal E}$. Substituting  
$\phi = \phi_0 + \varepsilon \phi_1$ into eq.(\ref{kink-energy}) we have,  
as one could expect, 
\begin{eqnarray}                  \label{kink-energy-1} 
{\cal E} &= M + \varepsilon^2 \displaystyle \int\limits_{-\infty}^{\infty} dx  
\displaystyle  \frac{1}{2}  
{\dot {\phi_1}}^2 + {\cal O}(\lambda)=M + \varepsilon ^2 
\displaystyle \frac {3 m^5} 
{\lambda {\sqrt 2}} t^2  + {\cal O}(\lambda) 
\nonumber\\ 
 &=M + \displaystyle \frac {M V^2}{2} +  {\cal O}(\lambda), 
\end{eqnarray}   
where $V = \varepsilon 3mt/\sqrt 2 = at$ is the kink velocity.  
 
Note, that the changing of the kink kinetic energy is equal to the changing of  
the potential energy of the field due to linear perturbation, because 
\begin{equation}  
\Delta {\cal V} = \varepsilon ~\frac{m^3}{\sqrt \lambda} \int dx~  
\left(\phi_0 + \varepsilon \phi_1 + \dots \right), 
\end{equation} 
 and, in the same second order, for the large $m t \gg 1$ we have 
\begin{equation} 
\Delta^{(2)} {\cal V} = - \varepsilon ^2 \frac{3 m^3}{2 {\sqrt {2}} \lambda}~ m^2 t^2  
\int \frac{dz}{\cosh^2 z} = - \varepsilon ^2\frac {3 m^5} 
{\lambda {\sqrt 2}} t^2 \equiv \frac {M V^2}{2}. 
\end{equation}

The second order correction  $\phi_2$ to the kink solution $\phi _0$ 
(\ref{kink})
 can be found from the next equation:  
\begin{equation}                   \label{second-equ} 
\biggl(\frac {d^2}{d t^2} + D^2\biggr) 
 \phi _2 + 3  \lambda \phi _0 \phi_1^2 = 0, 
\end{equation} 
where $D^2$ is the operator 
(\ref{second-der}) again and the first order correction  
$\phi _1$ is 
defined by (\ref{phi-one}).  
Suppose that at the moment $t=0$ all oscillation modes are excited, i.e. 
we take 
$a_1 = 1, a_k = 1$ for all values of $k$. Using again the  
expansion of   
$\phi _2$ on the eigenfunction 
$\eta _n (z)$ of the operator $D^2$ we write 
\begin{equation}                            \label{expan-two} 
\phi _2 = \sum _{n=0}^{\infty} \alpha_n(t) \eta _n (z) 
\end{equation} 
Substituting the expansion (\ref{expan-two}) in   
equation (\ref{second-equ}) 
after the projection onto the zero mode one can obtain: 
\begin{equation} 
\frac {4 }{3} {\ddot \alpha}_0 + 3  \lambda  
\int\limits_{-\infty}^{\infty}  dz 
~\phi_0 \phi_1 ^2 \eta _0 = 0 
\end{equation} 
where $\phi _1$ is defined by eq.(\ref{phi-one}). Thus, we have 
\begin{equation}                           \label{lambda-two} 
{\ddot \alpha}_0 + \frac{9m^3}{4 \sqrt \lambda } \int\limits_{-\infty}^{\infty} 
\frac {dz}{\cosh^2 z} \tanh z \biggl(- \frac {1}{2} +  
\frac{3}{4 \cosh^2 z} (1 - m^2t^2)  
+  \frac {\sinh z}{\cosh^2 z}e^{i \omega _1 t}  + 
\sum _{k=0}^{\infty}  e^{i \omega _k t}\eta_k \biggr)^2 = 0 
\end{equation}  
 
It is obvious, that if oscillation modes at the initial moment were not excited 
the correction to the zero mode would be equal to zero. Indeed,  
taking into account that 
\begin{equation} 
 \int\limits_{-\infty}^{\infty} dz \tanh z \frac {1}{\cosh^2 z}\biggl( 
1 + \frac {3}{\cosh^2 z} (1 - m^2t^2)+  
 \frac {9}{4\cosh^4 z} (1 - m^2t^2)^2\biggr) = 0, 
\end{equation} 
because the integrand is an odd function, we can write the solution of  
eq. (\ref{lambda-two}) as  
\begin{eqnarray} 
\alpha_0 & =e^{i \omega_1 t} \alpha (\omega_1, t) +  \displaystyle
\sum\limits_{k=0}^{\infty}  
e^{i \omega_1 t} \alpha (\omega_1, \omega_k) + \displaystyle 
\sum\limits_{k=0}^{\infty}  
e^{i \omega_k t}\alpha (k)  
\nonumber\\ 
& +  \displaystyle \sum\limits_{k=0}^{\infty}  e^{i \omega_k t}\alpha (k,t)  +  
\sum \limits_{k,k'=0}^{\infty}e^{i (\omega_k - \omega_k') t}\alpha (k, k')~~ 
 + ~~{\rm ~c.c.} 
\end{eqnarray} 
where the amplitudes of the oscillations of the kink are 
\begin{eqnarray}                  \label{cr-on} 
\alpha (\omega_1, t)   
= \frac {3\pi m}{64{\sqrt\lambda}} \left(-\omega_1^2 t^2 + 4i  
\omega_1 t + \frac {11}{2}\right); \quad 
\alpha (\omega_1,\omega_k) = \frac {9 \pi}{16 m {\sqrt \lambda}} 
\frac{ (\omega_k + \omega_1)^2  (1 + k^2)}{\cosh \frac{\pi k}{2}}; 
\end{eqnarray}
\begin{eqnarray}                   \label{cr-one} 
\alpha (k) 
=\frac {9 m i \pi}{16 {\sqrt\lambda}} \frac {k}{\sinh 
\frac{\pi k}{2}}; \quad 
\alpha (k,t) = \frac {9 m i \pi }{64 {\sqrt\lambda} }\left(\omega _k ^2 t^2  
+ 4 i  
\omega_k t - 6 - \frac {\omega _k^2}{m^2}\right)\frac {k^2}{\sinh
\frac{\pi k}{2}}; 
 \end{eqnarray}
and  
 \begin{eqnarray}                \label{cr-couple} 
\alpha (k, k') 
=\frac {9 i \pi  m}{8 {\sqrt \lambda}} \frac  
{\omega _k^2 - 
\omega _{k'}^2}{ (\omega _k - \omega _k')^2} \frac {1 +  
\frac {k^2 + {k'}^2}{4}} 
{~{\rm sh}\ \frac{\pi (k+ k')}{2}} 
\end{eqnarray} 
Here we take into account that the  integrals, which  
correspond to the  
transitions between different modes on the kink background  are: 
\begin{eqnarray}  
\int dz \tanh z \frac {\sinh z}{\cosh^4 z}\eta _k =& \displaystyle \frac {\pi} 
{4 m^4 } \frac {(\omega_k^2 - \omega_1^2)^2} {\cosh \frac{\pi k}{2}}  
(1 + k^2); \quad \displaystyle \int dz \tanh z \frac {\sinh z}{\cosh^4 z} 
=  \frac {\pi } {8};
\nonumber\\ 
\int dz \tanh z \frac {1}{\cosh^2 z}~\eta _k =& \displaystyle -i 
\frac{k \omega_k^2} 
{2m^2} \frac{\pi}{\sinh 
\frac{\pi k}{2}}; \quad \displaystyle \int dz \tanh z \frac {1} 
{\cosh^4 z}~\eta _k  = -i \frac{k^2\omega_k^4}{12 m^4}  
\frac{\pi}{\sinh \frac{\pi k}{2}}; 
\nonumber\\ 
 \int dz \tanh z \frac {1}{\cosh^2 z} 
 \eta _k \eta ^*_{k'} =& \displaystyle i\pi \frac{\left(\omega _k^2 - 
\omega _{k'}^2\right)\left(1 +  
\frac {k^2 + {k'}^2}{4}\right)}{2 m^2 \sinh \frac{\pi (k+ k')}{2}}; 
\quad  \displaystyle \int dz \tanh z \frac {\sinh z}{\cosh^6 z} 
=  \frac {\pi } {16}. \nonumber 
 \end{eqnarray} 
Note, that the amplitude of the excitation of the $k$-th continuum mode is  
exponentially supressed for $k \gg 1$. In fact, only $k, k' \leq 1 $ will  
contribute.

Thus, collecting all terms  together we have the next expression for the  
second order correction to the kink position: 
\begin{eqnarray}           \label{kink-corr-zero-2} 
\delta x^{(2)} = \displaystyle  \varepsilon ^2 \frac{{\sqrt{2 \lambda }}}{m^2} 
\alpha_0 =\displaystyle 
&-&  \displaystyle \varepsilon ^2 \frac {9 \pi }{4 {\sqrt 2}m} \biggl( 
 \sum _{k,k'=0}^{\infty}\sin (\omega_k - \omega_k') t ~\frac  
{\omega _k^2 -  
\omega _{k'}^2}{ (\omega _k - \omega _k')^2} \frac {2 +  
\frac {k^2 + {k'}^2}{2}} 
{\sinh \frac{\pi (k+ k')}{2}}  
\nonumber\\ 
&+ &\displaystyle \sum_{k=0}^{\infty} \cos (\omega_k - \omega_1)t~ 
\frac{1 + k^2}{\cosh \frac{\pi k}{2}} (\omega_k + \omega_1)^2  
\nonumber\\ 
 &-& \displaystyle \sum_{k=0}^{\infty}\sin \omega_k t \frac {k}{\sinh 
\frac{\pi k}{2}}   
+  \sum_{k=0}^{\infty}\sin  \omega_k t ~ \left( -\omega _k ^2 t^2   
 + 6 + \frac {\omega _k^2}{m^2}\right)\frac {k^2}{4 \sinh 
\frac{\pi k}{2}} 
\nonumber\\ 
&+& \displaystyle \sum_{k=0}^{\infty} \omega_k t \cos \omega_k t ~ 
\frac {k^2}{\sinh 
\frac{\pi k}{2}} + \frac{1}{3} \omega_1 t \sin \omega_1 t - \frac{1}{12} 
\left( 
\omega_1^2 t^2 - \frac{11}{2}\right) \cos \omega_1 t\biggr). 
\end{eqnarray}  
 
All these terms correspond to the oscillations of the kink by interaction 
with  
the vibrational modes. The first two  terms in eq.(\ref{kink-corr-zero-2})  
with time-independent amplitudes correspond to the scattering of phonons 
on kink 
(the first term) or the capture of phonon by kink  
(second term). The third  
term in (\ref{kink-corr-zero-2}) describes the   
oscillation of the kink due to interaction with the phonons   
created by the shift of the vacuum value of the scalar field (\ref{phi-vac}).  
Other terms contribute to the friction of the   
kink and contain the corrections to the velocity and acceleration of it.  
 
The energy of the kink interaction with phonons can be  
calculated from the second order correction $\delta x^{(2)} $  
eq.(\ref{kink-corr-zero-2}). Indeed, this correction is 
a sum of oscillations 
with the frequencies $\omega_k$ and different amplitudes $\delta x_k$.  
Thus, for the large time interval $t \gg 1$ 
the energy of each oscillation can be written as  
\begin{eqnarray} 
E_k = M\frac{\delta {x_k^{(2)}}^2\omega_k^2}{2} \approx M\frac
{\varepsilon^4 3^4 
\pi^2}{2^{13}} m^4t^4 \frac {k^4(4+k^2)^3}{\sinh^2 
\frac{\pi k}{2}} = MV^4 \frac{\pi^2}{2^{11}}\frac {k^4(4+k^2)^3}{\sinh^2 
\frac{\pi k}{2}}, 
\end{eqnarray} 
where $M$ and $V$ are the kink mass and velocity. 
 
Introducing the integration over momenta $k$ instead of  
sum one can obtain 
\begin{equation} 
\Delta E = \int\limits_{0}^{\infty}\frac{dk}{2\pi} E(k) = M V^4 
\frac{\pi}{2^{12}} 
\int\limits_{0}^{\infty}\frac{dk~ k^4 (4+k^2)^3}{\sinh^2 
\frac{\pi k}{2}} + MV^4 \frac{3 \pi^2}{2^9} 
\end{equation} 
where the last term is the contribution of the first vibration mode. 
 
Taking into account that  
$$\int\limits_{0}^{\infty}\frac{dk~ k^4 (4+k^2)^3}{\sinh^2 
\frac{\pi k}{2}} \approx 3\cdot  2^{12}  
$$  
we can estimate the total  
energy of interaction between the kink and phonons as   
\begin{equation}   
\delta E^{(2)}  \approx 3 \pi MV^4  
 \end{equation}   
that is much more than the second order relativistic correction (i.e.
$\frac{3}{8} MV^4$).  

In a simular way the second order corrections to the 
other kink modes can be obtained. Note, that among them there are terms, 
which describe
the production of  kink-antikink pair. Such a  
correction to the vibration mode  
$\eta _1$ was considered recently in \cite{Manton-4}. Here we consider these 
corrections supposing that at the moment $t=0$ all oscillation modes are  not 
excited, i.e. we take 
$a_1 = 0,  a_k = 0$ for all values of $k$. We have already seen that in this 
case 
the correction to the zero mode is equal to zero.
The correction of the second order to the vibrational mode $\eta _1 (z)$  
can be calculated after the projection of (\ref{second-equ}) onto this mode: 
\begin{equation}   
{\ddot \alpha}_1 + \omega_1^2 \alpha_1 + \frac{9m^3}{2 \sqrt \lambda } 
\int\limits_{-\infty}^{\infty} dz \tanh z  
\frac {\sinh z}{\cosh^2 z}\biggl(\displaystyle -\frac {1}{2} +  
\frac{3}{4 \cosh^2 z} (1 - m^2t^2) \biggr)^2 = 0. 
\end{equation}    
The solution of this equation  
is 
\begin{equation}                 
\alpha_1(t) =  \frac{m}{\sqrt{\lambda}} \left\{ b_1 e^{i\omega_1t} +  
\frac{3 \pi}{8} \left( -3 + \frac{1}{2} \omega_1^2 t^2 - \frac{1}{8} 
\omega_1^4 t^4 
\right) \right\},  
\end{equation} 
where $ b_1 = {\rm const}$  
and up to fluctuations corrections at  
the large $\omega_1 t \gg 1$ one has  
\begin{equation}              \label{lambda_2-1} 
\alpha_1 \approx  \frac{3 \pi}{2^6} \frac{m}{\sqrt  
\lambda }  \omega_1^4 t^4.  
\end{equation}  
 
The simular equation for the correction to the lowest mode of the continuum 
($k_0 = 0$)  
\begin{equation}       
{\ddot \alpha}_{k_{0}}  + 2m^2 \alpha_{k_{0}} + \frac{3m^3}{8 \pi \sqrt  
\lambda } \int\limits_{-\infty}^{\infty} dz \tanh z \displaystyle  
 (3  \tanh^2 z -1 ) 
\biggl( \displaystyle -\frac {1}{2} +  
\frac{3}{4 \cosh^2 z} (1 - m^2t^2) \biggr)^2 = 0 
\end{equation}    
obviously gives the trivial solution, because the integrand is an odd function. 
 
As for the correction to the other continuum modes, after the projection onto  
the k-$th$ mode, we have  
\begin{eqnarray}       
{\ddot \alpha}_{k} + \omega_k^2 \alpha_{k} &  
\nonumber\\ 
\displaystyle + \frac{3 m^3}{2 \pi \sqrt  
\lambda }\frac  {1}{(1+k^2)(4+k^2)} &\displaystyle   
\int\limits_{-\infty}^{\infty} dz~  
\tanh z  ~ \eta_k ~\biggl(\displaystyle -\frac {1}{2} +  
\frac{3}{4 \cosh^2 z} (1 - m^2t^2) \biggr)^2 = 0 
\end{eqnarray}    
 
Taking again the limit $m t \gg 1$ we obtain   
\begin{eqnarray}       
{\ddot \alpha}_{k} + \omega_k^2 \alpha_{k} =  \frac{9 i ~m^3}{2^9 
\sqrt\lambda }~  
\frac {4+k^2}{1+k^2}~ \frac{k^2}{\sinh  
\frac{\pi k}{2}} ~m^4 t^4 
\end{eqnarray}    
The solution of this equation is  
\begin{equation}                \label{lambda_2-k} 
\alpha_k \approx \frac{9 i~ m}{2^8 \sqrt\lambda } ~\frac {k^2}{1 + k^2} ~\frac  
{ m^4 t^4}{\sinh \frac{\pi k}{2}} + b_k e^{i\omega_kt} 
\end{equation} 
and the oscillation correction here can be dropped again. 
 
Thus, collecting the contributions of eqs. (\ref{lambda_2-1}), 
(\ref{lambda_2-k}), we obtain the time-dependent part of the 
second order correction to the kink classical field: 
\begin{equation}   \label{phi-two} 
\phi _2 = \frac {9 m}{2^8 \sqrt\lambda} m^4 t^4 \left\{ 3 \pi \eta_1 +  
i \sum _{k\not= 0} \frac{k^2}{1+k^2}~ \frac {1}{\sinh \frac{\pi k}{2}} ~\eta_k 
\right\}. 
\end{equation}
The next order correction to the energy can be found when we substitute  
$\phi = \phi_0 + \varepsilon \phi_1 + \varepsilon^2 \phi_2$  
into eq.(\ref{kink-energy}). Using the orthogonality relations, we see that  
$\phi_1$ is orthogonal to $\phi_2$ and the non-zero is only the  
fourth-order correction  
 
\begin{equation}           \label{kink-energy-2} 
\delta^{(4)} {\cal E} = \varepsilon^4 \int\limits_{-\infty}^{\infty} dx   
\frac{1}{2} {\dot {\phi_2}}^2 = \frac{3}{ 2^{10}} \left( 3 \pi^2 - 
\frac{1}{4} \int dk~ \frac{4+k^2}{1+k^2}~ \frac{k^4}{\sinh^2 \frac{\pi k}{2}} 
\right) M V^4 m^2 t^2. 
\end{equation} 
 
The first term in the roots corresponds to the kink mass changing due to  
its bounding with the $\eta_1$ mode.      
Numerical calculation of the integral over $k$ in the roots gives the  
value $\approx 5.406$, i.e.   
the second term in (\ref{kink-energy-2}) is, in two order,  
smaller then first one.  
This term corresponds to the correction to energy of the configuration 
connected 
with the continuum modes excitation.  
 
\section{Interaction of the t'Hooft-Polyakov monopole with external 
homogeneous magnetic field.} 
The t'Hooft-Polyakov monopole \cite{Hooft,Pol} is a well-known static 
solution of the nonlinear Yang-Mills-Higgs field equations.  
Though  
considerable progress has been achieved in the last two decades, there are 
still open problems concerning the dynamical properties of monopoles.  The 
most known results were obtained in the Bogomolny-Prasad-Sommerfield (BPS) 
limit \cite{BPS} where the monopole dynamics changes drastically due to the 
masslessness of the scalar field.  A calculation of the static force 
between two monopoles \cite{Manton-1} and of the light scattering by a 
monopole \cite{BakLee} were based on an ansatz for the time dependence of 
the field which was just a replacement ${\vec r} \to {\vec r} - \frac{1}{2} 
{\vec w}t^2$ for the monopole position ${\vec r}$. That corresponded 
already to monopoles moving with a constant acceleration ${\vec w}$.  In 
Ref.\cite{Manton-1} as well as in the followed papers \cite{Gold-Wali,LOR}, 
the interaction between monopoles was considered in the region outside the 
monopole core where the Yang-Mills fields obey the free field equations. 
However, it is reasonable to expect a distortion of the core of the 
t'Hooft-Polyakov monopole and a Bremstrahlung of both vector and scalar 
fields if an initially static monopole configuration is accelerated by an 
external field. 
 
In this note we describe a consistent perturbative consideration of this 
idea. Only the lowest-order result is presented here. It  
shows the 
monopole acceleration expected by the Newton law. The next-to-leading 
corrections will decrease ${\vec a}$ because of the radiation. 
 
Let us consider the $(3+1)$D $SU (2)$  Yang-Mills-Higgs model 
specified by the Lagrangian: 
\begin{equation}                                     \label{Lagrang-YM} 
L = \frac{1}{4} {F_{\mu \nu}^a} {F^{\mu \nu}} ^a + \frac {1}{2} D_{\mu }
\phi ^a 
D^{\mu }\phi ^a + 
\frac {\lambda}{4} (\phi ^2 - a^2)^2, 
\end{equation} 
where 
$ 
F_{\mu \nu}^a = \partial _{\mu} A_{\nu}^a -  \partial _{\nu} A_{\mu} ^a  + 
e \varepsilon _{abc} 
 A_{\mu} ^b  A_{\nu} ^c;\quad D_{\mu }\phi ^a = \partial _{\mu} \phi ^a + 
e \varepsilon _{abc} A_{\mu} ^b \phi  ^c.$ 
 
In order to consider the interaction of this configuration with an external 
magnetic field let use the analogy with above considered example of the  
2D kink motion. As in the $\phi ^4$ model concidered above 
we introduce the Lagrangian of interaction 
which is linear on scalar field: 
\begin{equation}        \label{inter-mono} 
L_{int} = 
\frac{1}{2a}\varepsilon _{kmn} F_{km}^a \phi ^a 
 H_n^{(ext)}, 
\end{equation} 
where $ H_n^{(ext)}$ is an external homogeneous constant  
magnetic field which is supposed 
to be small (i.e. $ \mid H_n^{(ext)} \mid \ll a^2 $). Taking into 
account the definition of 
the $U(1)$ electromagnetic subgroup one can see that this  
term describes direct 
interaction between the magnetic field of the  
t'Hooft-Polyakov monopole and the external   
magnetic field. 
 
The field equations take the form: 
\begin{equation}                                \label{field-eq-YM} 
D^{\mu} F_{\mu \nu}^a =  e \varepsilon _{abc} \phi ^b D_{\nu } 
 \phi ^c + {\it F}_{\nu} ^a; 
\qquad 
D^{\mu} D_{\mu} \phi ^a = \lambda (\phi ^2 - a^2) \phi ^a + {\it 
F}^a. 
\end{equation} 
where the  last terms represent the external force acting on the 
configuration. They read 
\begin{equation}                   \label{external} 
{\it F}_{0} ^a = 0;\quad 
{\it F}_{n} ^a = - \frac{1}{a} \varepsilon _{mnc} D_m \phi ^a 
H_c^{(ext)}; \quad 
 {\it F} ^a =  \frac{1}{2a} \varepsilon _{mnc}  F_{mn}^a H_c^{(ext)}. 
\end{equation} 
 
The key point of our approach is to treat the excitation of the zero modes 
of the monopole as a non-trivial time-dependent translation of the whole 
configuration. The amplitude of this excitation can be calculated from 
the field equation (\ref{field-eq-YM}). To this end, in  
analogy with the above 
considered case of the 2D $\lambda \phi^4$ model, we expand the fields 
$A_{\mu}^a, \phi^a$: 
$A_{\mu}^a = (A_{\mu}^a)_0 +  a_{\mu}^a + \dots;~ 
\phi ^a = (\phi ^a)_0 +  \chi^a + \dots $. 
The zero 
order approximation gives the classical equations 
\begin{equation}                                      \label{zero-equ-YM} 
D^{\mu } F_{\mu \nu}^a =  e  \varepsilon _{abc} \phi ^b D_{\nu } \phi ^c; 
\quad 
D^{\mu }D_{\mu } \phi ^a = \lambda (\phi ^2 - a^2) \phi ^a 
\end{equation} 
with the t'Hooft--Polyakov monopole solution \cite{Hooft,Pol}: 
\begin{equation}                          \label{t'Hooft-Polyakov} 
A_0 ^a = 0; \quad A_k^a =  \varepsilon _{akc} \frac{r^c}{e r^2} 
\left(1 - 
K(\xi)\right);\quad 
\phi ^a = \frac {r^a}{e r^2} H (\xi), \quad {\rm where}~ \xi = aer. 
\end{equation} 
  
Note that in the BPS limit \cite{BPS} ($\lambda \to 0$), one has instead of 
the field equations 
(\ref{zero-equ-YM}) a simpler first order equation 
$ 
D_m\phi^a = \displaystyle\frac{1}{2} \varepsilon_{knm}F_{kn}^a \equiv 
B_m^a. 
$ 
Then field equations (\ref{field-eq-YM}) take the form 
\begin{equation} 
\left(D_{m} - \frac {1}{a} H_m^{(ext)}\right)F_{mn}^a =  e  \varepsilon _{abc} 
\phi ^b D_{n} \phi ^c;\quad 
D_{m}\left(D_{m} - \frac {1}{a} H_m^{(ext)}\right)\phi ^a = 0. 
\end{equation} 
This is exactly the Manton's equation for a slowly accelerated monopole 
in a weak uniform magnetic field \cite{Manton-1}. 
 
Let us consider corrections to the 't Hooft--Polyakov solution. 
To the first order, they can be found from equations 
\begin{eqnarray}                   \label{first-equ-YM} 
\left( -\frac{d^2}{dt^2} + {\cal D}^2_{(A)}\right) a_n^a & \equiv & 
D^{\mu } D_{\mu } a_n^a  - e^2 [(\phi^a )^2 \delta _{ab} - 
\phi ^a \phi ^b ] a_n^b + e \varepsilon_{abc} a_m^b F_{mn}^c\\ 
&=& 2 e \varepsilon_{abc} \chi^b D_n \phi ^c + {{\it F}_{n}^a}^{(l)};    
\nonumber\\ 
\left( -\frac{d^2}{dt^2} + {\cal D}^2_{(\phi)}\right) \chi ^a 
&\equiv & D^{\mu } D_{\mu } \chi ^a - e^2 
[(\phi^a )^2 \delta _{ab} - 
\phi ^a \phi ^b ]\chi ^b - \lambda [2 \phi^a \phi ^b + ((\phi^a )^2 
- a^2) \delta _{ab}] 
\chi ^b 
\nonumber\\ 
 &=& - 2 e \varepsilon_{abc} a_n^b D_n \phi ^c + {{\it F}^a}^{(l)}, 
\nonumber 
\end{eqnarray} 
where the superscript $(l)$ corresponds to the direction of the external 
field and the background gauge $D_{\mu} a_{\mu} ^a + e \varepsilon _{abc} 
\phi ^b \chi ^c = 0$ is used. In the matrix notations these equations can 
be rewritten in the form 
\begin{equation}                           \label{first-equ-mon} 
\left(-\frac {d^2}{d t ^2}   + {\cal D}^2 \right)f^a   = 
{{\cal F}^a}^{(l)}, 
\end{equation} 
where 
$f^a ({\bf r}, t) = \left(\begin{array}{c} 
a_n^a({\bf r}, t)\\ 
\chi ^a({\bf r}, t)\end{array} \right), \quad 
{{\cal F}^a}^{(l)} = \left(\begin{array}{c} 
{{\it F}_{n}^a}^{(l)}\\ 
{{\it F}^a}^{(l)}\end{array} \right), 
$ 
and ${\cal D}^2$ is the matrix obtained after two functional differentiations 
of the action with respect to the fields $A_{\mu}^a, \phi^a$ 
$$ 
{\cal D}^2 f^a = \left( 
\begin{array}{cc} 
{\cal D}^2_{(A)} a_n^a   & -2 e \varepsilon_{abc} \chi^b D_n \phi ^c \\[3pt] 
2 e \varepsilon_{abc} a_n^b D_n \phi ^c  & {\cal D}^2_{(\phi)}\chi ^a 
\end{array} \right). 
$$ 
 
We seek for the solution of eq. (\ref{first-equ-mon}) in the form of 
an expansion 
\begin{equation}                                              \label{fexpan} 
f^a ({\bf r}, t)=\sum\limits_{i=0}^{\infty}C_i(t)\zeta^a_i{\bf r}) 
\end{equation} 
on the complete set of eigenfunctions $\zeta^a_i{\bf r})$ of the operator 
${\cal D}^2$. These eigenfunctions consist of a vector- and a scalar 
component: 
$\zeta^a_i{\bf r}) = \left(\begin{array}{c} 
\eta _n^a({\bf r})\\ 
\eta ^a({\bf r})\end{array} \right)$ 
describing the fluctuations of the corresponding fields on the monopole 
backgroung \cite{Rossi}. 
 
Substituting of expansion (\ref{fexpan}) into eq.(\ref{first-equ-YM}) 
results in the following system of equations for coefficients $C_i(t)$: 
\begin{eqnarray}                            \label{expan-YM-modes} 
\sum _{i = 0}^{\infty} \left({\ddot C}_i + \Omega_i^2 C_i \right) 
{\eta_n^a({\bf r})}_i - 2 e \varepsilon_{abc} \chi^b D_n \phi ^c 
=   {{\it F}_n^a({\bf r})}^{(l)}; 
\nonumber\\ 
\sum _{i = 0}^{\infty} \left({\ddot C}_i + \omega_i^2 C_i \right) 
{\eta ^a ({\bf r})}_i + 2 e \varepsilon_{abc} a_n^b D_n \phi ^c 
=  {{\it F}^a ({\bf r})}^{(l)}, 
\end{eqnarray} 
 
Let us consider a correction to the monopole solution, contributed by the 
excitation of the zero modes 
${\zeta^a_0}^{(k)} = \left(\begin{array}{c} 
{{\eta_n^a({\bf r})}}_0^{(k)}\\ 
{{\eta^a({\bf r})}}_0^{(k)}\end{array} \right)$ where \cite{Rossi} 
\begin{eqnarray}                   \label{modeYM-zero} 
{\eta_n^a({\bf r})}_0^{(k)} = F_{kn}^a = 
\partial _k  A_n^a - D_n A_k^a; \quad 
{\eta ^a({\bf r})}_0^{(k)} = D_k \phi^a = \partial _k \phi ^a - e 
\varepsilon_{abc} \phi^b A_k^c. 
\end{eqnarray} 
Here the index $k$ corresponds to the translation  in the direction ${\hat 
r}_k$.  These modes are normalized in such a way that makes $C_0$ in 
expansion (\ref{fexpan}) equal to the displacement of the monopole. 
Note that these zero modes coincide with the pure translational quasi-zero 
modes of the vector and scalar fields ~ ${{\widetilde \eta_n^a} ({\bf 
r})}^{(k)} = \partial _k  A_n^a; ~ {{\widetilde \eta^a}({\bf r}) }^{(k)} = 
\partial _k \phi ^a $  up to a gauge transformation with a special choice 
of the parameter which is just the gauge potential $A_k^a$ itself. 
 
A projection of eq.(\ref{expan-YM-modes}) onto 
the zero modes gives the following equation: 
\begin{equation}                              \label{modeYM} 
{\ddot C}_0 (N_v^2 + N_s^2) = \int d^3 x {{\it F}_n^a({\bf r})}^{(l)} 
{{\eta _n^a 
({\bf r})}_0}^{(k)} + \int d^3 x  {{\it F}^a({\bf r})}^{(l)} 
{\eta ^a({\bf r})}_0^{(k)}, 
\end{equation} 
(note that the non-diagonal terms cancel), 
where the normalization factors of the zero modes are 
\begin{eqnarray}                                    \label{orto-zero} 
N_v^2 = \int d^3 x \left[{\eta_n^a({\bf r})}_0^{(k)} \right]^2 
 = \int d^3 x \left({ F_{kn}^a }\right)^2; 
\quad 
N_s^2 = \int d^3 x \left[{\eta ^a({\bf r})}_0^{(k)} \right]^2 
= \int   d^3 x \left({D_k \phi^a }\right)^2 . 
\end{eqnarray} 
There is a very simple relation between the 
monopole zero modes 
normalization factors and the mass of the monopole $M$: 
\begin{eqnarray} 
N_v^2 + N_s^2 &=& \int d^3 x \left\{ 
\left({{\eta_n^a}_0^{(k)}}\right)^2 + 
 \left({{\eta^a}_0^{(k)}}\right)^2 \right\} = \int d^3 x \left\{ \left( 
F_{kn}^a \right)^2 +  \left(D_k \phi^a \right)^2  \right\} 
\nonumber\\ 
 &=&\frac {1}{2} \int d^3 x  \left\{ \left( 
F_{kn}^a \right)^2 +  \left(D_k \phi^a \right)^2  \right\} + 
\frac {1}{2} \int d^3 x ~V[\phi] = M . 
\end{eqnarray} 
The integrals in the r.h.s. of eq.(\ref{modeYM}) are calculable as well. We 
find 
\begin{equation} 
\int d^3 x {{\it F}_n^a({\bf r})}^{(l)} 
{{\eta _n^a({\bf r})}_0}^{(k)} =  \frac {2 H_k^{(ext)}}{3a}\int d^3 x 
D_m \phi^a B_m^a = \frac {2}{3} g H_k^{(ext)}, 
\end{equation} 
where we take into account the de\-fi\-ni\-tion of the mono\-pole 
charge ${\mbox{$g = \int  d^3 xD_m \phi^a B_m^a = 4 \pi/e$}}$. 
In a similar way 
\begin{equation}                                              \label{norm2} 
\int d^3 x 
 {F^a({\bf r})}^{(l)} 
{\eta ^a({\bf r})}_0^{(k)} = \frac {1}{a} \int d^3 x B_l^a H_l^{(ext)} 
D_k \phi^a = \frac {1}{3} g H_k^{(ext)}. 
\end{equation} 
We would like to stress that the only zero modes along the external field 
direction $H_k$ are excited. 
 
A substitution of eq.(\ref{orto-zero})--(\ref{norm2}) into eq.(\ref{modeYM}) 
gives the final equation for the monopole zero mode evolution in the form 
$M{\ddot C}_0 =  g{\cal H}$. Thus we obtain  
\begin{equation} 
{\ddot C}_0 =  \frac{g{\cal H}}{M}. 
\end{equation} 
Thus the monopole moves under external force along the external field 
direction 
with a constant acceleration $w = g{\cal H}/{M}$. This corresponds to the 
classical Newton formula with the Lorentz force $F = g{\cal H} = Mw$.  The 
radiative corrections to this relation are given by the next orders of 
perturbation theory. 
 
Note that the excitation of the monopole zero modes (\ref{modeYM-zero}) 
leads not only to the displacement of the solution but also to its time 
dependent gauge transformation $ {\cal U} = \exp\left\{-\frac{w_k 
t^2}{2e}A_k^a T^a\right\}$: 
\begin{eqnarray*} 
A_{n}^a({\vec r}) &\to& A_{n}^a({\vec r}) + 
C_0 {\eta _n^a({\vec r})}_0^{(k)} = 
{\cal U}~  A_{n}^a({\vec r} - {\vec w}t^2/2) ~{\cal U}^{-1} +  {\cal U} 
~ \partial _n {\cal U}^{-1};\\ 
A_{0}^a({\bf r}) &\to&  {\cal U}~ \partial _0 ~{\cal U}^{-1} = 
-\frac{1}{e} w_k t A_k^a ({\bf r});\\ 
\phi^a ({\vec r}) &\to& \phi^a  + 
C_0 {\eta^a({\vec r})}_0^{(k)} = {\cal U}~ 
\phi^a ({\vec r} - {\vec w}t^2/2)~ {\cal U}^{-1}. 
\end{eqnarray*} 
Thus, the monopole motion in coordinate space is connected with its 
rotation in the isotopic space with a constant angular acceleration.  It 
follows that the electric field of the accelerated monopole is $E_n^a 
= \frac{1}{a} H_k^{(ext)} F_{nk}^a t$ as it should be. 
 
Note that the Lagrangian of interaction  (\ref{inter-mono}) is linear in 
the scalar field as well as its analog  (\ref{Lagrang})  
in 2D $\lambda \phi^4$ model.  
If one considers it as a correction to the Higgs potential,  
then it also lifts the degeneracy of the vacuum (see 
fig.3). Indeed, on the electromagnetic  
asymptotics the potential 
\begin{equation} 
V[\phi] = \frac {\lambda}{4} (\phi ^2 - a^2)^2 + \frac{1}{a} 
B_{n}^a \phi ^a H_n^{(ext)} 
\end{equation} 
has the minimum 
\begin{equation} 
\phi^a_{min} = a {\hat r}^a \left( 1 + e \frac{{\hat r}_n }{r^2} 
\frac{H_n^{(ext)}} 
{m_s^2 m_v^2} \right) 
\end{equation} 
where the standard notation for the masses of scalar $(m_s^2 = 2 \lambda a^2)$ 
and 
vector $(m_v^2 = e^2 a^2)$ particles was used.  
Thus the minimum of $ V[\phi]$ lies in the 
direction of the external magnetic field. 
 
Let us justify the form of the interaction Lagrangian (\ref{inter-mono}). 
A direct interaction between the monopole configuration and an external 
homogeneous non-Abelian gauge field  do not affect the zero modes.  Indeed, 
let us introduce this interaction as 
\begin{equation} 
L_{int} = B_k^a {H^a_k }^{(ext)}. 
\end{equation} 
\noindent where ${H^a_k }^{(ext)}$ has only one non-zero component 
${H^3_z}^{(ext)}~ = {\rm const}$. 
In this case the field equations modified as 
\begin{equation}                                  \label{field-eq-YM-3} 
D_{m} F_{mn}^a =  e \varepsilon _{abc} \phi ^b D_n 
\phi ^c  + \varepsilon _{nmc} D_m {H^a_c }^{(ext)}; 
\end{equation} 
$$ 
D_m D_m \phi ^a = \lambda (\phi ^2 - a^2) \phi ^a. 
$$ 
The external force 
${\cal F}_n^a =\varepsilon _{nmc} D_m {{H^a_c} }^{(ext)}$ 
is now acting on the vector field only.  It is orthogonal to the 
monopole zero modes, and, therefore, it does not excite them. Indeed, 
\begin{eqnarray*} 
\lefteqn{\int d^3 x {\cal F}_n^a 
{{\eta _n^a({\bf r})}_0}^{(k)} = \int d^3 x \varepsilon _{nmc} 
D_m {{H^a}_c }^{(ext)} F_{kn}^a }\\ 
=&{}\displaystyle\int d^3 x (1-K) \displaystyle 
\frac{r^a}{er^4}\left\{{H^k_a }^{(ext)} (1-K^2) - 
\xi {H^a_k }^{(ext)} \frac {dK}{d \xi} +  {\hat r}^k{\hat r}^m 
{H^a_m }^{(ext)} \left(1 - K^2 +  \xi  \frac {dK}{d \xi} 
\right) \right\} . 
\end{eqnarray*} 
This is equal to zero as an integral of an odd function. 
 
Let us note that we could use the gauge invariant Lagrangian of the 
electromagnetic interaction instead of eq.(\ref{inter-mono}): 
\begin{equation}                             \label{inter-mono-2} 
L_{int} = B_k H_k^{(ext)} = \frac{1}{2a}\varepsilon _{kmn} 
\left( F_{km}^a \phi ^a 
- \frac {1}{ea^2}\varepsilon _{abc}\phi ^a  D_k \phi ^b   D_m \phi ^c 
\right) H_n^{(ext)}. 
\end{equation} 
However, the additional term in eq.(\ref{inter-mono-2}) has no effect 
because the additional external force on the configuration  
which appears in 
r.h.s. of eq.(\ref{first-equ-mon}) is also orthogonal to the monopole 
zero modes 
(\ref{modeYM-zero}). Thus, the monopole interaction with the external 
electromagnetic field is determined only by the first term in 
eq.(\ref{inter-mono-2}) which is eq.(\ref{inter-mono}).   
The physical 
meaning of this result is quite obvious, because the second term in the 
gauge invariant definition of the electromagnetic field strength  in 
(\ref{inter-mono-2}) corresponds to the Dirac monopole string in the Abelian 
theory. 
 
Finally, consider an interaction between two widely separated t'Hooft--Polyakov 
mono\-poles with charges $g_1$ and $g_2$.  Let us suppose that the first 
monopole is placed at the origine while the second one is at the large 
distance $R \gg r_c$, $r_c$ stands for the core size.  To the leading order 
in $r_c/R$, the field of the second monopole can be considered as a 
homogeneous external field acting on the first one.  Thus, the 
electromagnetic part of the interaction is defined by the Lagrangian 
(\ref{inter-mono}) as before, where now $H_k^{(ext)} = - g_2 R_k/R^3$ and 
the first monopole will move with a constant acceleration 
$ 
w_k = g_1 H_k^{(ext)}/M = g_1 g_2 R_k/ 
(M R^3) = F_k/M 
$. 
This corresponds to the classical Coulomb force between the monopoles: 
$F_k = g_1 g_2 {R_k}/{R^3}$. 
 
As it has been noted in \cite{Manton-1,Gold-Wali,LOR}, this simple picture 
is not valid in the BPS limit.  Indeed, in this case the scalar field is 
also massless and there are long-range forces mediated by both the scalar 
and the electromagnetic fields of the monopoles. The scalar interaction is 
given by the term 
$ L_{int}' = D_m {\phi^a}^{(1)} D_m {\phi^a}^{(2)} $ 
in addition to the pure electromagnetic one (\ref{inter-mono}). 
As the Bogomolny condition gives 
$D_m \phi^a = B_m^a = \phi^a B_m$ for both monopoles, 
it takes the form 
\begin{equation} 
L_{int}' = \frac{1}{a}\phi^a 
D_m {\phi^a}^{(1)} 
{H_k}^{(ext)} =  \frac{1}{2a}\varepsilon _{mnk} F_{mn}^a \phi ^a 
H_k^{(ext)} . 
\end{equation} 
Thus, the total Lagrangian of interaction between two BPS monopoles is just 
double eq.(\ref{inter-mono}) in the case of the monopole-antimonopole 
configuration and equal to zero in case of the monopole-monopole (or 
antimonopole-antimonopole) configuration.

\section*{Acknowledgments} 
 
We gratefully acknowledge the support of Alexander von 
Humboldt Foundation and by Royal Society/NATO Fellowship (Ya.S.).
One of us (Ya.S.) thanks Professor 
R.~Seiler for his support and encouragement.  Ya.S. is also very grateful 
to D.~Diakonov, N.Manton, Per Osland and V.~Petrov and for very  helpful 
discussions  and would like to thank the University of Bergen where a part 
of this work  was  done,  for  hospitality  and support.  
This research was completed while one of us (Ya.S.) was visiting the  
International Centre for Theoretical Physics Trieste, Italy.

\newpage
%%%%%%%%%%%%%%%%%%%%%%%%%%%%%%%%%%%%%%%%%%%%%%%%%%%%%%%%%%%%%%%%%%%%%%%%
%%%%%%%%%%%%%%%%%%%%%%%%%%%%%%%%%%%%%%%%%%%%%%%%%%%%%%%%%%%%%%%%%%%%%%%% 
\begin{figure}[htb] 
\begin{center} 
\setlength{\unitlength}{1cm} 
\begin{picture}(15,7.) 
\put(-0.5,-0.5) 
{\mbox{\epsfysize=6.4cm\epsffile{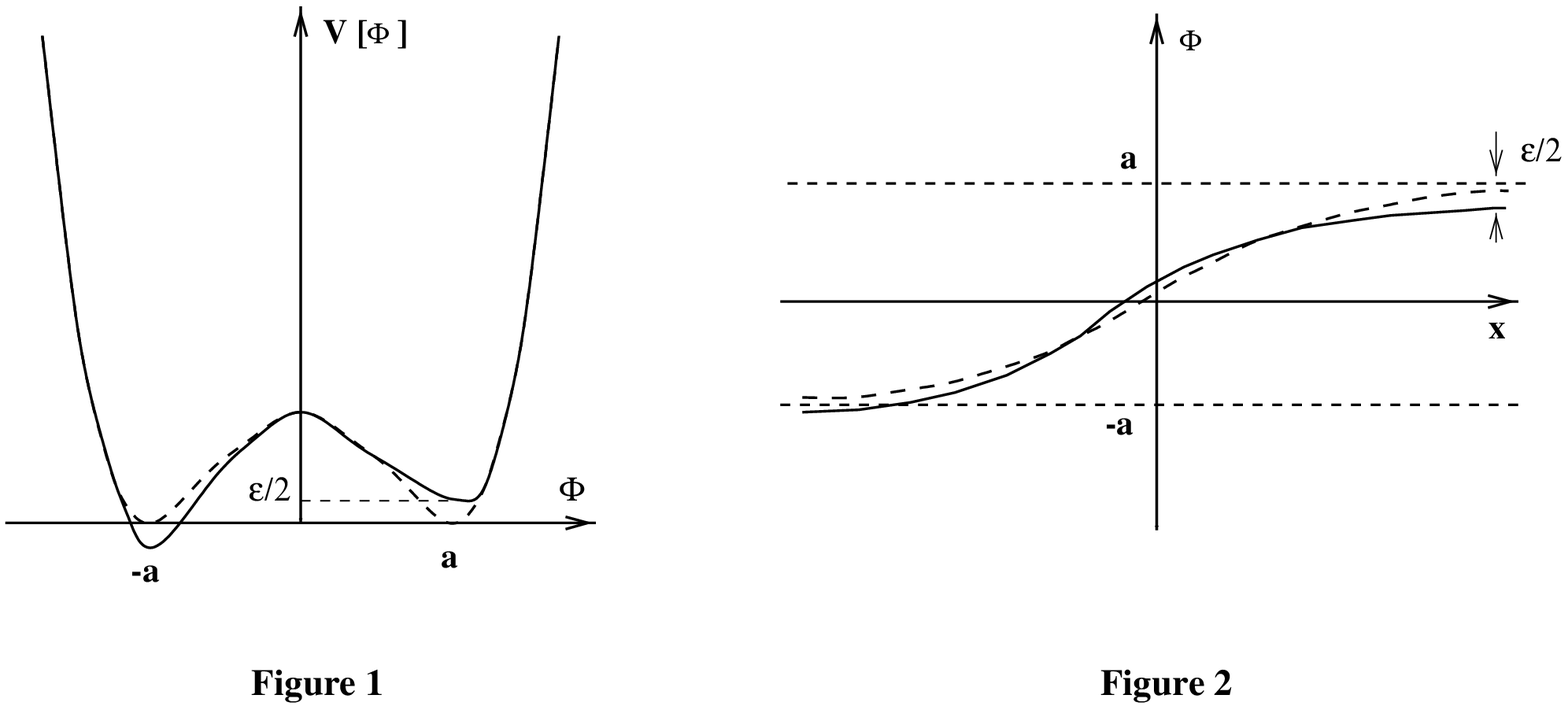}}} 
\end{picture} 
%\caption{~} 
\end{center} 
\end{figure} 
%%%%%%%%%%%%%%%%%%%%%%%%%%%%%%%%%%%%%%%%%%%%%%%%%%%%%%%%%%%%%%%%%%%%%%%% 
~
\bigskip
~

%%%%%%%%%%%%%%%%%%%%%%%%%%%%%%%%%%%%%%%%%%%%%%%%%%%%%%%%%%%%%%%%%%%%%%%% 
\begin{figure}[htb] 
\begin{center} 
\setlength{\unitlength}{1cm} 
\begin{picture}(15,7.) 
\put(2.7,-0.5) 
{\mbox{ \psfig{figure=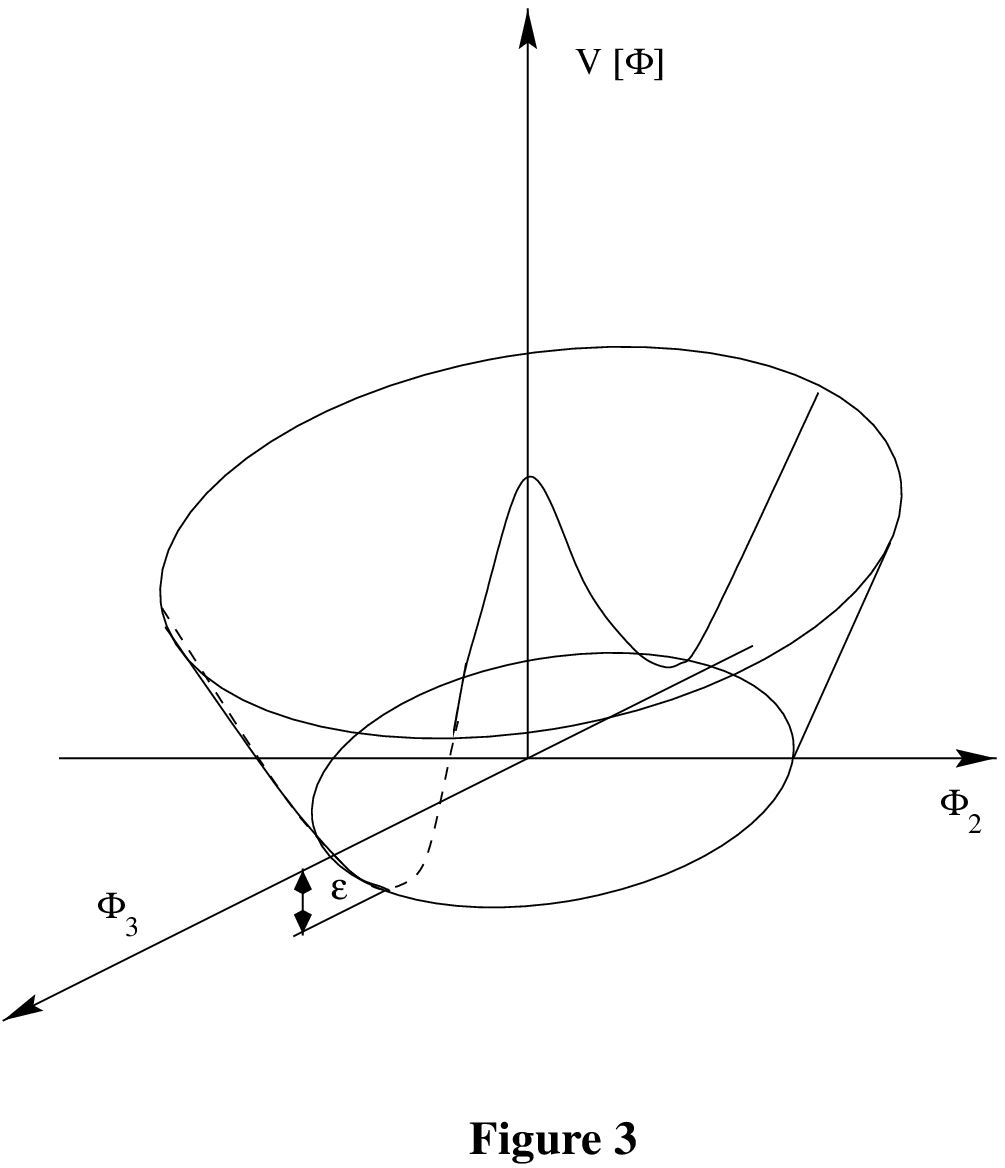,height=8.4cm}}} 
\end{picture} 
%\caption{~}
\end{center} 
\end{figure} 
%%%%%%%%%%%%%%%%%%%%%%%%%%%%%%%%%%%%%%%%%%%%%%%%%%%%%%%%%%%%%%%%%%%%%%%% 
 
\end{document}